\documentclass[type1rest,genTeX,reqno]{nrc2}

\usepackage{amsmath,amssymb}
\usepackage{graphicx}
\usepackage[figuresright]{rotating}
\usepackage{bm}
\usepackage{rotating}

\numberby{equation}{section}

\begin{document}

\title{A Tale of Two Horizons}

\author{Sean Stotyn}
\address{Department of Physics and Astronomy, University of British Columbia, Vancouver, British Columbia, Canada, V6T 1Z1}
\correspond{sstotyn@phas.ubc.ca}
\shortauthor{S. Stotyn}
                   

\begin{abstract}
I revisit the fate of coinciding horizons and the volume between them in the extremal limit of spherically symmetric black holes in four spacetime dimensions, focusing on the Schwarzschild de Sitter black hole for concreteness.  The two Killing horizons in the limit spacetime that are traditionally identified with the limiting event horizons of the non-extremal black hole are shown to instead be generated by an enhanced symmetry of the near horizon geometry (NHG). This dismantles the interpretation of the 4-volume between the horizons remaining finite in the extremal limit. The NHG is reinterpreted as a tangent spacetime to the degenerate black hole horizon, and geometrical objects, such as Killing vectors and Killing horizons, are carefully mapped between the bulk and the NHG. The implications for extremal black hole entropy are then discussed.
\end{abstract}

\maketitle

\section{Introduction}

The laws of black hole mechanics bare a striking resemblance to the laws of thermodynamics. Indeed, this similarity becomes solidified when quantum effects are taken into account: a black hole horizon emits thermal radiation at a characteristic temperature\cite{Hawking:1974sw}, and the information in the interior of the horizon (i.e. the entropy) is proportional to the spatial cross-sectional area of the event horizon\cite{Bekenstein:1973ur}. The source of this entropy largely remains a mystery due to the lack of a full quantum theory of gravity that would unambiguously identify the various microstates. However, there has been some progress in understanding black hole entropy using holography as a tool to count those microstates.

This was first carried out for extremal black holes\cite{Strominger:1996sh} and was later extended to the near-extremal case\cite{Horowitz:1996fn}. Extremal black holes are very different beasts from their non-extremal and even near-extremal counterparts. The causal structure of an extremal black hole is very different from that of a non-extremal black hole: in general, the extremal limit is not a Hausdorff limit. However, there is a sense of consistency since the laws of black hole mechanics forbid a non-extremal black hole from becoming extremal in a finite number of steps. This is because extremal black holes are at zero temperature. 

In contrast to most thermodynamic systems the nonzero horizon area suggests a nonzero entropy, although there is debate about how extremal black hole entropy is defined. The string theory and holography calculations put forth in \cite{Strominger:1996sh} predict the Bekenstein-Hawking entropy $S_{BH}=\frac{A_H}{4G}$ where $A_H$ is the horizon area and $G$ is Newton's constant, yet semiclassical considerations suggest the gravitational entropy is due to a nontrivial spacetime topology and should vanish for extremal black holes\cite{Hawking:1994ii,Teitelboim:1994az}. Similarly, the Wald entropy of an extremal black hole should vanish because the Killing horizon is not bifurcate\cite{Wald:1993nt}. Then again, recent work identifies the entanglement entropy between causally disconnected regions of spacetime with black hole entropy\cite{Myers:2013lva}, implying a nonzero entropy for extremal black holes. As is pointed out in \cite{Carroll:2009maa}, clearly not all of these methods calculate the same quantity, so how extremal black hole entropy is defined is still up for debate. It is my aim here to demonstrate why methods that rely on the global properties of the near horizon geometry (NHG) are poor choices for this definition.

If one takes semiclassical gravity seriously, then the Euclidean instantons of extremal black holes must also contribute to the path integral. One such case of historical importance is the Euclidean Nariai instanton, which is typically identified as the extremal Schwarzschild de Sitter (SdS) black hole. It is well-known that non-extremal SdS has no corresponding complete instanton, since there is no choice of the periodicity of Euclidean time that can simultaneously remove both conical singularities. This is physically interpreted as the black hole and cosmological horizons being at different temperatures. When considering the semiclassical contribution of SdS in \cite{Ginsparg:1982rs}, Ginsparg and Perry showed that although Euclidean SdS is not an instanton, it has a finite action that smoothly limits from the $S^4$ instanton of de Sitter space to the $S^2\times S^2$ Nariai instanton, further entrenching the identification of Nariai as extremal SdS. 

Examining the static patch between the horizons as the limit to extremality is taken, the authors of \cite{Ginsparg:1982rs} were led to the conclusion that the 4-volume between the horizons remains finite in the extremal limit; i.e. the two merging horizons would never overlap. This implies a non-zero temperature for extremal SdS, with the two horizons being at a common temperature, instead of a degenerate horizon at zero temperature. In addition to this interpretation not making intuitive sense (an extremal black hole implies a degenerate horizon), there are a number of subtleties associated with spacetime limits that must be carefully dealt with, in particular the hereditary and nonhereditary properties of spacetimes. In this paper, I show that the Nariai solution is in fact the NHG of extremal SdS, which is furthermore to be thought of as a tangent spacetime to the degenerate horizon. When viewed in this light, the non-degenerate Nariai horizons are actually generated by a Killing vector of the enhanced spacetime symmetry in the NHG.

This paper is outlined as follows. The Ginsparg-Perry limit of SdS is outlined in section 2 with a focus on where and why the standard interpretation arises. In section 3, I review the salient features of Geroch's seminal paper on the limits of spacetimes, in particular what happens to the Killing vectors when a limit is taken. I use these results in section 4 to show how to map geometrical data from the extremal spacetime into its NHG. I finish in section 5 with a discussion of these results and how they can be extended to other types of extremal black holes. In particular, I discuss the relation of these results to the unresolved issue of extremal black hole entropy.

\section{The Coinciding-Horizon Limit}
\label{sec:GP}

Let us begin by reviewing the Ginsparg-Perry limiting procedure found in Ref. \cite{Ginsparg:1982rs}, which concerns the Euclidean sector of a Schwarzschild de Sitter (SdS) black hole. The following procedure can be carried out in the Lorentzian sector and can be extended to isolated two-horizon black holes with different asymptotic structures,\footnote{These statements will be explicitly demonstrated in a future paper.} but for concreteness we will consider Euclidean SdS here. The metric is given locally by
\begin{eqnarray}
&&ds^2=f(r)d\tau^2+\frac{dr^2}{f(r)}+r^2d\Omega^2, \label{eq:EucSdS} \\
&&f(r)=1-\frac{2M}{r}-\frac13\Lambda r^2, \nonumber
\end{eqnarray}
where $d\Omega^2$ is the metric on a unit 2-sphere. There are two singularities at $r_+$ and $r_c$ leaving $r$ restricted to the range $r_+<r<r_c$; these singularities correspond to the black hole and the cosmological horizons respectively of the Lorentzian solution. One of these singularities can be removed by choosing an appropriate periodicity of $\tau$, but both cannot be simultaneously removed for an arbitrary choice of $M$ and $\Lambda$.

If $9M^2\Lambda=1$, the two singularities overlap $r_+=r_c$, seemingly leaving an ill-defined range for the $r$-coordinate. However, consider setting $9M^2\Lambda=1-3\epsilon^2$ so that $r_+=r_0(1-\epsilon)$ and $r_c=r_0(1+\epsilon)$, with $r_0=\frac{1}{\sqrt{\Lambda}}\left(1-\frac16\epsilon^2\right)$, and defining a new radial coordinate $\rho$ and a time coordinate $\psi$ via
\begin{equation}
r=r_0+\epsilon\rho, \>\>\>\>\>\>\>\>\>\>\>\>\>\>\> \tau=\frac{\psi}{\epsilon}. \label{eq:GPcoordtrans}
\end{equation}
Note that in this coordinatization, $r_c$ and $r_+$ correspond to $\rho=\pm r_0$ respectively. The metric (\ref{eq:EucSdS}) takes the form
\begin{equation}
ds^2=\left(1-\frac{\rho^2}{r_0^2}\right)d\psi^2+\frac{d\rho^2}{1-\frac{\rho^2}{r_0^2}}+r_0^2d\Omega^2+{\mathcal O}\big(\epsilon^2\big), \label{eq:EucNariai}
\end{equation}
which is regular in the $\epsilon\rightarrow 0$ limit. In the strict limit, it is the metric on $S^2\times S^2$ where each two-sphere has radius $r_0$; this is the Euclidean Nariai solution.  

\begin{figure}[t]
\centering
\includegraphics[width=0.9\linewidth]{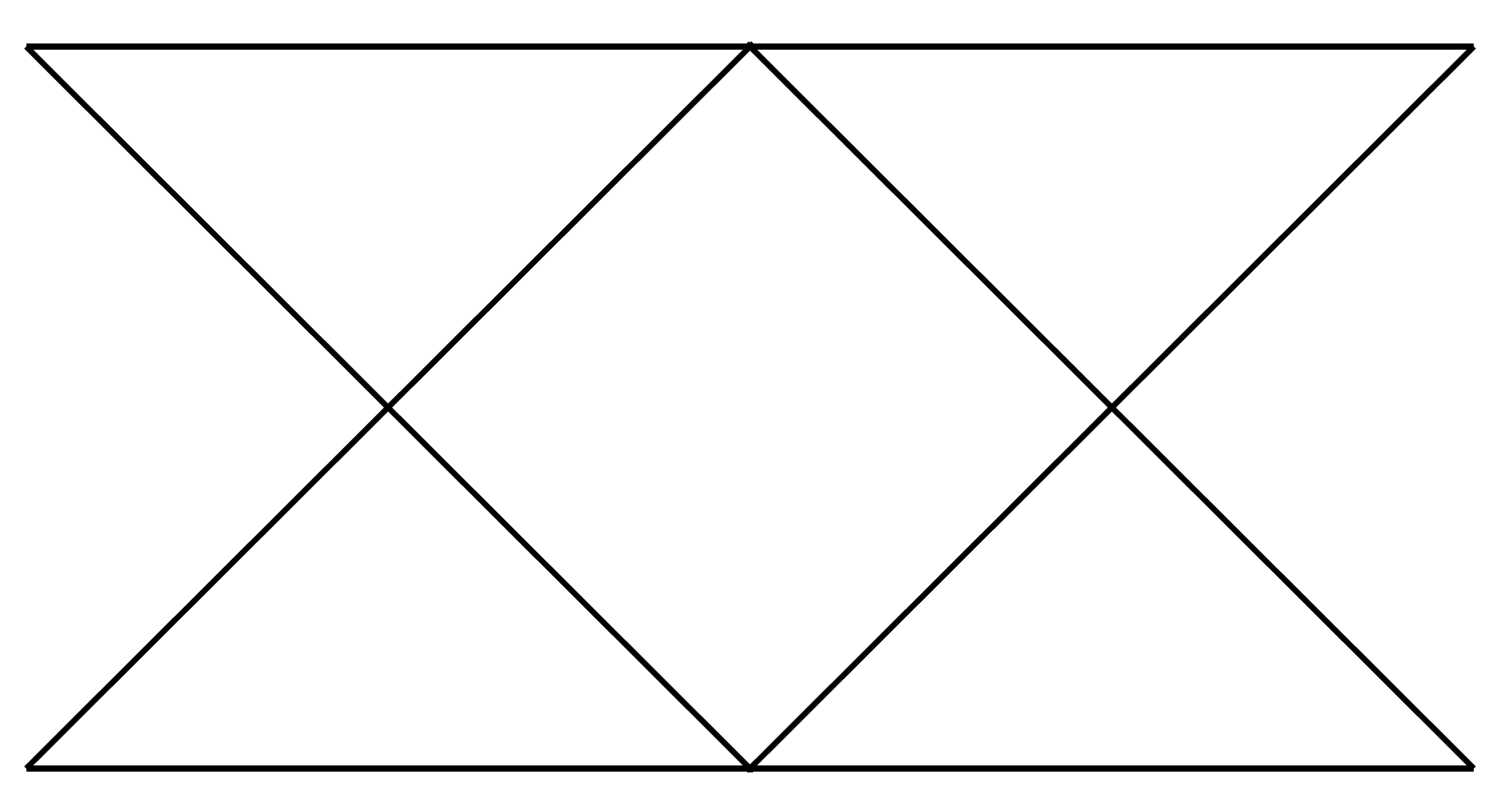}
\put(-230,60){$\cdots$}
\put(-5,60){$\cdots$}
\put(-163,75){\begin{turn}{45}$\rho=-r_0$\end{turn}}
\put(-90,98){\begin{turn}{-45}$\rho=r_0$\end{turn}}
\put(-159,42){\begin{turn}{-45}$\rho=-r_0$\end{turn}}
\put(-91,20){\begin{turn}{45}$\rho=r_0$\end{turn}}
\put(-180,119){\small $\rho=-\infty$}
\put(-70,119){\small $\rho=\infty$}
\put(-134,59){static patch}
\caption{The conformal diagram for the dS$_2$ part of the Nariai spacetime. The coordinates of Eq. (\ref{eq:Nariai}), cover the static patch indicated.}
\label{fig:StaticNariai}
\end{figure}
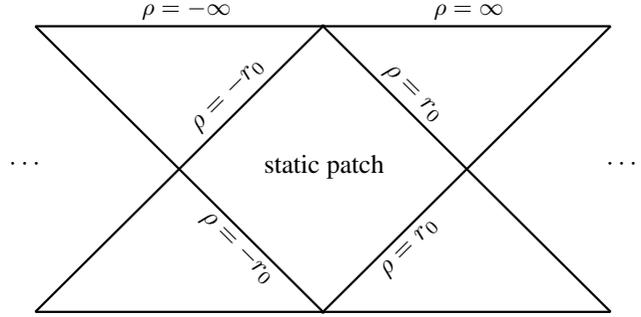

The $\epsilon\rightarrow0$ limit of the metric (\ref{eq:EucNariai}) is globally regular and clearly has finite 4-volume. The singularities at $r_c$ and $r_+$ of (\ref{eq:EucSdS}) remain separated at $\rho=\pm r_0$ and are both removed by choosing $\psi$ to have period $2\pi r_0$. Because this limit is performed in the Euclidean sector, necessitating the existence of a static region, the standard interpretation of this result is that the 4-volume of the static patch between the horizons of SdS remains finite in the extremal limit. Thus, the Nariai solution is often identified as the extremal SdS black hole. We will see shortly, however, that Nariai, whose Lorentzian sector is dS$_2\times S^2$ given by
\begin{equation}
ds^2=-\left(1-\frac{\rho^2}{r_0^2}\right)dt^2+\frac{d\rho^2}{1-\frac{\rho^2}{r_0^2}}+r_0^2d\Omega^2, \label{eq:Nariai}
\end{equation}
is instead the near-horizon geometry of extremal SdS, which has no static regions. The conformal diagram of the dS$_2$ part of Nariai is displayed in Fig. \ref{fig:StaticNariai}, showing the static patch covered by (\ref{eq:Nariai}) and the two horizons located at $\rho=\pm r_0$.

The Nariai instanton's contribution to the partition function implies a tunnelling rate from de Sitter space to extremal SdS. However, the stability of Euclidean Nariai against transverse trace-free symmetric tensor perturbations $h_{ab}$ was also examined in \cite{Ginsparg:1982rs}, where it was found that the spectrum of the differential operator $G$ defined by $G_{abcd}h^{ab}=-\Box h_{cd}-2R_{acbd}h^{ab}$ has a single negative eigenvalue. This Euclidean negative mode causes the degenerate horizon to split into an outer and inner horizon, ending in the black hole evaporating on a time scale much shorter than the timescale of nucleation. This analysis suggests that extremal SdS always has a quantum instability toward decaying to a non-extremal SdS black hole, but this analysis rests on the underlying assumption that Nariai is the same as the extremal SdS black hole, since it is the former that contains the instability. Since Nariai will be shown to be the NHG of extremal SdS, it is unclear how the global properties of Nariai (i.e. the spectrum of the operator $G$) impacts the stability of the extremal SdS horizon.

\section{Limits of Spacetimes}
\label{sec:limits}

As outlined in Geroch's seminal paper on spacetime limits \cite{Geroch:1969ca}, which properties are preserved in a limit is a nontrivial matter. To begin with, ``the" limit of a spacetime is an ill-conceived notion because many independent limits may exist. For instance, starting with the Schwarzschild metric with $M=\epsilon^{-3}$
\begin{equation}
ds^2=-\left(1-\frac{2}{\epsilon^3 r}\right)dt^2+\frac{dr^2}{\left(1-\frac{2}{\epsilon^3 r}\right)}+r^2d\Omega^2,
\end{equation}
there is clearly no limit as $\epsilon\rightarrow0$. However, the following two diffeomorphisms each yield a valid $\epsilon\rightarrow0$ limit:
\begin{eqnarray}
&&r=\tilde r/\epsilon, \>\> t=\epsilon\tau, \>\> \theta=\epsilon\rho, \\
&&ds^2\rightarrow\frac{2}{\tilde r}dt^2-\frac{\tilde r}{2}d\tilde r^2+\tilde r^2(d\rho^2+\rho^2 d\phi^2) \nonumber\\
&&r=x-\epsilon^{-4}, \>\> \theta=\epsilon^4\rho. \\
&&ds^2\rightarrow-dt^2+dx^2+d\rho^2+\rho^2 d\phi^2. \nonumber
\end{eqnarray}
The former is a non-flat Kasner metric, while the latter is the flat Minkowski metric. Evidently curvature invariants are not preserved in spacetime limits as this simple example explicitly demonstrates.

In fact, there are certain properties of the parent spacetime which are preserved in limits -- so called hereditary properties -- and others that are not. Since the horizons being considered in this paper are all Killing horizons, let's examine the hereditary properties of Killing vectors. To do so, first consider a one-dimensional family of manifolds $(M_\epsilon,g_{ab}(\epsilon))$. Out of these, we construct a five-dimensional manifold $\overline{M}$ with signature $(0,-,+,+,+)$ with the $(M_\epsilon,g_{ab}(\epsilon))$ as leaves of the foliation. It was shown in \cite{Geroch:1969ca} that conformal transformations of $\overline M$ also obey the various limits, so there exists a conformal completion $\widetilde M=\overline{M}\cup\partial\overline{M}$. The boundary $\partial\overline{M}$ is called the limit space and corresponds to $(M_0,g_{ab}(\epsilon=0))$. The necessity of the degenerate direction in $\overline{M}$ is due to there being several independent valid limits of $\epsilon\rightarrow0$.

Next, if $K_\epsilon$ denotes the set of Killing vectors on $(M_\epsilon,g_{ab}(\epsilon))$, then it can be shown that $\dim K_0\ge\dim K_\epsilon$. In particular, this states that symmetries cannot be lost when taking limits, but symmetries can be gained. A crucial feature of this analysis is that  $K_\epsilon$ is not required to smoothly map to $K_0$ in the $\epsilon\rightarrow0$ limit, but rather all that is required is for it to limit to an accumulation space. 

This may seem counterintuitive since the geometry must limit smoothly to $(M_0,g_{ab}(\epsilon=0))$ and Killing vectors are isometries of the metric. However, this subtle point is responsible for the interpretation of the Ginsparg-Perry limit above. Indeed, examining the coordinate transformation (\ref{eq:GPcoordtrans}), it is evident that the timelike Killing vector transforms as $\partial_t\rightarrow\epsilon\partial_\tau$, which is singular in the $\epsilon\rightarrow0$ limit. This implies that the Killing horizons generated by $\partial_t$ in the parent spacetime ($r_+$ and $r_c$) do not smoothly map to the Killing horizons generated by $\partial_\tau$ in the limit spacetime ($\pm r_0$). Therefore, there is no meaningful sense in which the 4-volume between the horizons remains finite in the extremal limit.

\section{Mapping of Geometric Objects}

\begin{figure}[t]
\centering
\includegraphics[width=\linewidth]{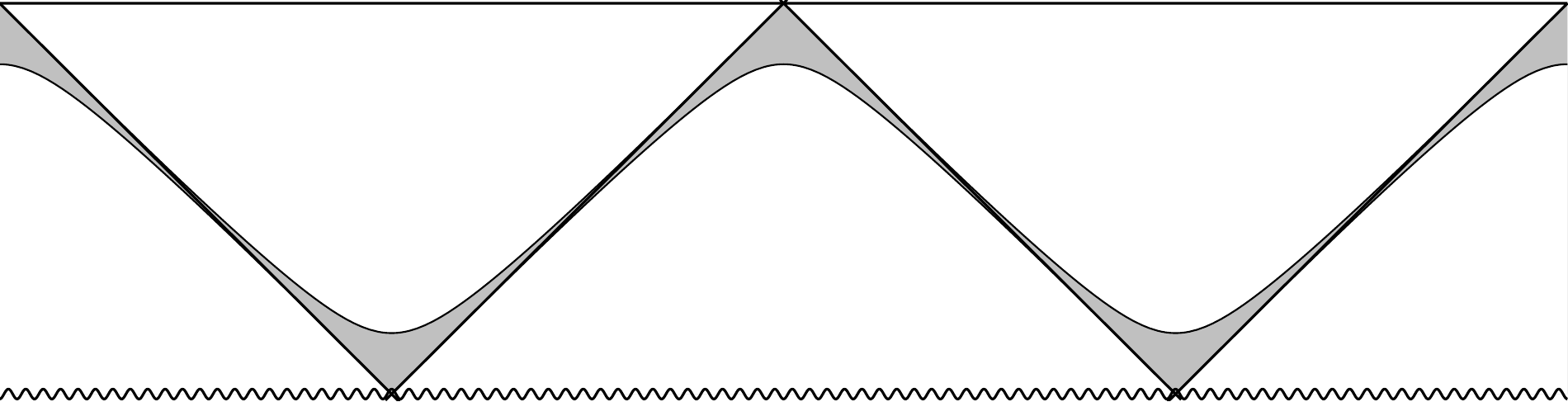}
\put(-168,28){\begin{turn}{45}$\tau=r_0$\end{turn}}
\put(-106,49){\begin{turn}{-45}$\tau=r_0$\end{turn}}
\put(-137,4){\small $\tau=0$}
\put(-200,66){\small $\tau=\infty$}
\put(-76,66){\small $\tau=\infty$}
\put(-127,27){I}
\put(-66,37){II}
\put(-189,37){II}
\caption{The conformal diagram for extremal SdS. Regions I and II are separately covered by (\ref{eq:ExtremalSdS}), but the chart is not regular across $\cal H$ located at $\tau=r_0$. The shaded regions represent the small neighbourhood $U({\cal H})$.}
 \label{fig:XSdS}
\end{figure}

The extremal SdS solution can be obtained from the canonical limit, $9M^2\Lambda\rightarrow1$, of SdS. This limit is well-defined and the metric is given locally by
\begin{eqnarray}
&&ds^2=-\frac{d\tau^2}{V(\tau)}+V(\tau)d\sigma^2+\tau^2d\Omega^2, \label{eq:ExtremalSdS} \\
&&V(\tau)=\left(\frac{\tau+2r_0}{3\tau}\right)\left(\frac{\tau}{r_0}-1\right)^2, \nonumber
\end{eqnarray}
where $r_0=\Lambda^{-1/2}$. Note that there is no static patch anywhere, hence no static timelike Killing vector. However, there is a degenerate Killing horizon $\cal H$ located at $\tau=r_0$ generated by $\partial_\sigma$, which is spacelike everywhere except at $r_0$ where it is null. Furthermore, as can be seen from its conformal diagram in Fig. \ref{fig:XSdS}, this spacetime is not a black hole but rather it is a cosmology with an initial singularity evolving asymptotically toward de Sitter.

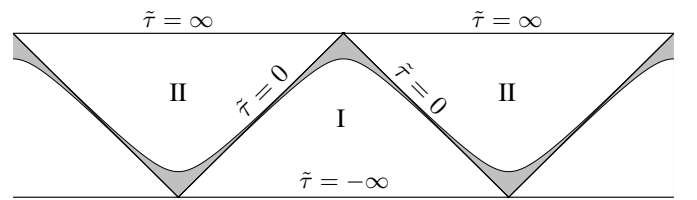
\begin{figure}[b] 
\centering
\includegraphics[width=\linewidth]{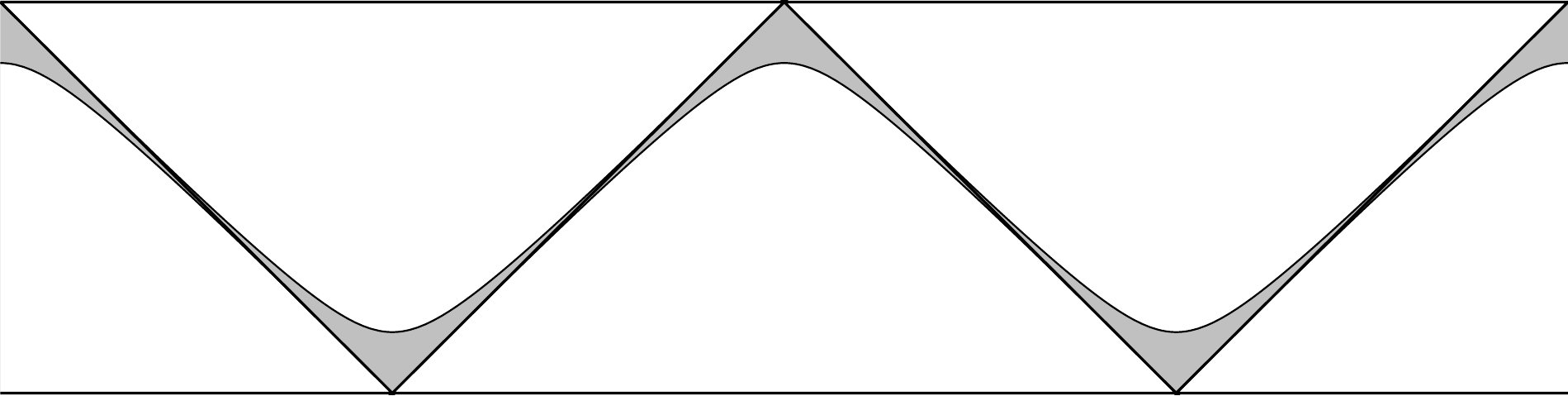} 
\put(-166,30){\begin{turn}{45}$\tilde\tau=0$\end{turn}}
\put(-106,48){\begin{turn}{-45}$\tilde\tau=0$\end{turn}}
\put(-141,4){\small $\tilde\tau=-\infty$}
\put(-200,65){\small $\tilde\tau=\infty$}
\put(-76,65){\small $\tilde\tau=\infty$}
\put(-127,27){I}
\put(-66,37){II}
\put(-190,37){II}
\caption{The conformal diagram for dS$_2$. Regions I and II are separately parametrized by the coordinates $\tau$ and $\sigma$. The shaded regions correspond to $U({\cal H}_N)$.}
\label{fig:dS2xS2}
\end{figure}

Using Gaussian null coordinates in the neighbourhood of an extremal horizon, it has been shown that any degenerate horizon admits a NHG that satisfies the same equations of motion (see Ref. \cite{Kunduri:2013gce} for an extensive review). Inspired by this, I examine the NHG by expanding the metric (\ref{eq:ExtremalSdS}) about $\tau = r_0$, and keeping only the lowest order terms:
\begin{equation}
ds^2\approx-\frac{r_0^2}{\tilde\tau^2}d\tilde\tau^2+\frac{\tilde\tau^2}{r_0^2}d\sigma^2+r_0^2 d\Omega^2, \label{eq:NHG}
\end{equation}
where $\tilde\tau\equiv \tau-r_0$.
It is now easily verified that this metric is dS$_2\times S^2$, i.e. the Nariai solution. Note that this procedure does not guarantee that (\ref{eq:NHG}) satisfies the equations of motion. Instead, this procedure recovers the NHG and provides a map $\phi:~U({\cal H})\rightarrow U({\cal H}_N)$ between the bulk and the NHG, where $U(p)$ is a small open neighbourhood containing $p$ and ${\cal H}_N$ is defined by the image of $\phi$ acting on $\cal H$.

To connect (\ref{eq:NHG}) with the discussion in section \ref{sec:GP}, the following coordinate transformation is employed
\begin{equation}
\tilde\tau = -r_0e^{-t/r_0}\sqrt{1-\frac{\rho^2}{r_0^2}}, \>\>\>\>\>\>\>\> \sigma=\frac{\rho e^{t/r_0}}{\sqrt{1-\frac{\rho^2}{r_0^2}}}. \label{eq:CoordTrans}
\end{equation}
$\cal H$ is located at $\tilde\tau=0$, which means ${\cal H}_N$ is located at $\rho=\pm r_0$ and $t=+\infty$. Eq. (\ref{eq:CoordTrans}) puts the NHG metric into the standard Nariai chart
\begin{equation}
ds^2=-\left(1-\frac{\rho^2}{r_0^2}\right)dt^2+\frac{d\rho^2}{1-\frac{\rho^2}{r_0^2}}+r_0^2 d\Omega^2, \label{eq:NariaiTwo}
\end{equation}
which was obtained in the Ginsparg-Perry limit in Eq. (\ref{eq:Nariai}). 

\begin{figure}[t] \label{fig:NariaiNbhd}
\centering
\includegraphics[width=0.9\linewidth]{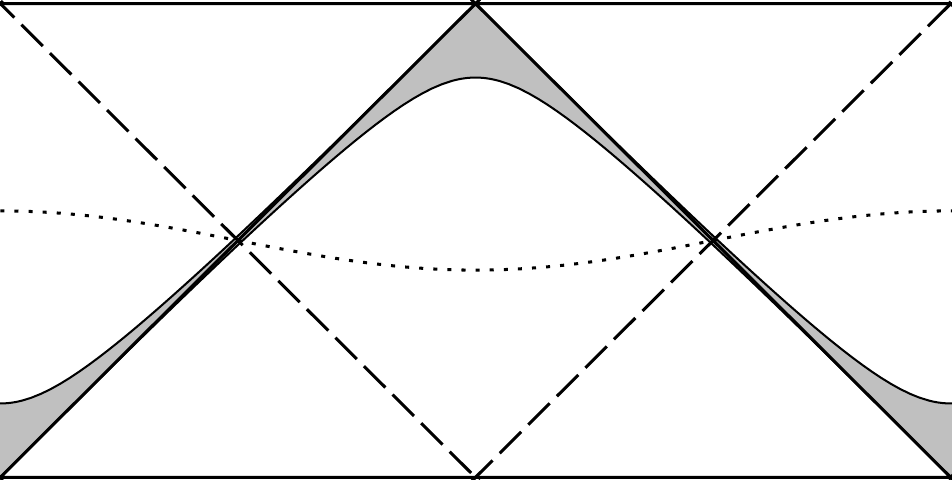}
\put(-132,60){\small static patch}
\put(-167,70){\begin{turn}{45}${\cal H}_N,~{\cal H}_t^+$\end{turn}}
\put(-92,97){\begin{turn}{-45}${\cal H}_N,~{\cal H}_t^+$\end{turn}}
\put(-156,30){\begin{turn}{-45}${\cal H}_t^-$\end{turn}}
\put(-86,20){\begin{turn}{45}${\cal H}_t^-$\end{turn}}
\caption{The conformal diagram for dS$_2$. The diamond-shaped static patch is covered by coordinates $t$ and $\rho$. The dashed lines are ${\cal H}_t^-$, the solid lines are ${\cal H}_N$ and ${\cal H}_t^+$, the dotted line is the $t=0$ slice, and the shaded regions correspond to $U({\cal H}_N)$.}
\end{figure}
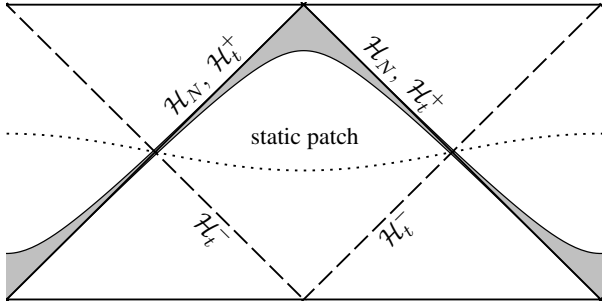

The explicit form of the map $\phi$ is given by Eq. (\ref{eq:CoordTrans}), allowing for an unambiguous mapping of geometrical data. Recall that $\cal H$ is generated by the Killing vector $K=\partial_\sigma$, which transforms according to
\begin{equation}
K\rightarrow \frac{-\rho e^{-t/r_0}}{r_0\sqrt{1-\frac{\rho^2}{r_0^2}}}\partial_t+\sqrt{1-\frac{\rho^2}{r_0^2}}e^{-t/r_0}\partial_\rho. \label{eq:KV}
\end{equation}
One can explicitly verify that $K$ is a Killing vector of the Nariai metric, (\ref{eq:NariaiTwo}). It has norm $K^2=e^{-2t/r_0}\left(1-\frac{\rho^2}{r_0^2}\right)$ and is everywhere spacelike except at $\rho=\pm r_0$, $t=+\infty$ where there is a degenerate root; $t=-\infty$ is not contained in ${\cal H}_N$. The horizons ${\cal H}_t$ generated by $\partial_t$, on the other hand, are located at $\rho=\pm r_0$, with the future horizons ${\cal H}_t^+$ located at $t=+\infty$ and the past horizons ${\cal H}_t^-$ located at $t=-\infty$. Hence, ${\cal H}_N\cap{\cal H}_t^+$ is non-empty, whereas ${\cal H}_N\cap{\cal H}_t^-$ is empty. The horizon structure is shown in Fig. \ref{fig:NariaiNbhd}.

Finally, applying the inverse map $\phi^{-1}$, it is easily verified that $\partial_t\rightarrow-\big(\frac{\tau}{r_0}-1\big)\partial_\tau+\frac{\sigma}{r_0}\partial_\sigma$, which is not a Killing vector of the bulk spacetime (\ref{eq:ExtremalSdS}). Evidently, the Killing vector $\partial_t$ is a generator of the increased spacetime symmetry group in the near horizon region. In fact, this must be true by a very simple argument: the NHG must contain the conformal group, yet the extremal SdS spacetime possesses no static timelike Killing symmetries. Therefore the time translation symmetry must be due to a generator of the enhanced symmetry group.

\section{Discussion}

It is well-established that spacetimes with degenerate horizons admit NHGs as solutions to the same equations of motion. Furthermore, these NHGs possess an increased spacetime symmetry group that includes (a subgroup of) the conformal group. In this paper, I have shown that the generator of time translations arising from this increased symmetry group is responsible for the interpretation of the 4-volume remaining finite between merging horizons. To illustrate the problem, I have taken the canonical extremal limit of the SdS black hole and expanded the metric about the degenerate horizon ${\cal H}$, yielding the Nariai metric in unconventional coordinates. The NHG was then reinterpreted as a tangent spacetime to $\cal H$; it is only valid in a small neighbourhood $U({\cal H}_N)$, where ${\cal H}_N$ is the image of ${\cal H}$ mapped into the NHG.

The analysis presented here can easily be generalised to black holes with merging inner and outer horizons, such as the extremal limit of Reissner-Nordstr{\o}m. The details of this and more general cases is deferred to a future paper. Focusing on the geometry between the two horizons, there is an analogous Ginsparg-Perry limit (although necessarily with Lorentzian signature) which naively leads to the conclusion that the 4-volume in this region remains finite in the extremal limit\cite{Carroll:2009maa}. As is done in this paper, one may take the canonical extremal limit of the Reissner-Nordstr{\o}m black hole and expand the metric about the degenerate horizon $\cal H$: in this case one obtains AdS$_2\times S^2$ as the NHG and there exists a map $\tilde\phi: U({\cal H})\rightarrow U({\cal H}_N)$ which brings the NHG metric into global AdS coordinates. This again demonstrates that the NHG is a tangent spacetime to the degenerate horizon and hence the inverse map $\tilde\phi^{-1}$ does not apply globally.

The implications of this for extremal black hole entropy are far reaching. From a string theory perspective, extremal black holes (particularly supersymmetric black holes) are local vacua of the theory. They are thought of as the intersection of stacks of $D$-branes, where the dynamics of the $D$-branes decouple from the bulk and the branes act like a free gas\cite{Maldacena:1997re}. This is known as the decoupling limit and involves sending the Planck length to zero while simultaneously performing a diffeomorphism on the near-extremal black hole. For spherical black holes in $d$ dimensions, the result is a NHG metric of the form AdS$_2\times S^{d-2}$. Next, the AdS/CFT correspondence is used to identify the CFT microstates on the boundary with the microstates of the extremal horizon, leading to the Bekenstein-Hawking entropy. Alternatively, the microstates can be obtained by the CFT on the worldsheet of the effective string defined along the brane intersections of a higher dimensional black string. However, in this case the central charge and the oscillator levels of the affine currents are all defined asymptotically. Both of these methods rely on global properties of the NHG, so their connection to the microstates of the bulk spacetime are suspect. Furthermore, time translations of the CFT are with respect to an enhanced symmetry in the NHG, not time translations in the bulk.

Considering the Wald entropy \cite{Wald:1993nt} in the NHG presents an interesting dilemma for extremal black hole entropy. From inspection of Fig. \ref{fig:NariaiNbhd}, ${\cal H}_t$ is a bifurcate Killing horizon and hence ought to yield a nonzero entropy, while ${\cal H}_N$ is clearly not bifurcate and hence ought to yield zero entropy. It appears that even in the NHG itself there are conflicting notions of how entropy should be defined. Given that a point $p'$ in the NHG only maps to a point $p$ in the bulk iff $p'\in U({\cal H}_N)$, I propose defining extremal black hole entropy with respect to ${\cal H}_N$. However, the aim of this paper is not to settle the issue of extremal black hole entropy, but rather to explicitly highlight some subtle and often under-appreciated properties of NHGs. 

First, they are tangent spacetimes to the degenerate horizons of extremal black holes. Second, time translations in the NHG are generated by one of the increased symmetries, implying notions of energy and temperature in the NHG are disconnected from those of the bulk. Third, the extremal black hole horizon maps to a degenerate Killing horizon that partially overlaps the non-degenerate Killing horizon generated by the increased symmetry. These subtle points force careful and deliberate thought about the thermodynamics of extremal black holes, in particular how significantly they differ from their non-extremal (even near-extremal) counterparts. I leave further exploration of these and related issues, including extensions to other types of black holes and black branes, for future work.

\section*{Acknowledgements}

The author would like to thank Kristin Schleich and Don Witt for helpful discussions. This work was made possible by funds supplied by the Natural Sciences and Engineering Research Council of Canada.

\bibliographystyle{plain}

\end{document}